# Effects of Zn on the grain boundary properties of $La_{2-x}Sr_xCu_{1-y}Zn_yO_4$ superconductors

**S.H. Naqib\* and R.S. Islam**

Department of Physics, Rajshahi University, Rajshahi-6205, Bangladesh

**Abstract**

The properties of the grain boundaries (GBs) are of significant importance in high-$T_c$ cuprates. Most large scale applications of cuprate superconductors involve usage of sintered compounds. The critical current density and the ability to trap high magnetic flux inside the sample depend largely on the quality of the GBs. Zn has the ability to pin vortices but it also degrades superconductivity. In this study we have investigated the effect of Zn impurity on the intergrain coupling properties in high-quality $La_{2-x}Sr_xCu_{1-y}Zn_yO_4$ sintered samples with different hole concentrations, $p$ ($\equiv x$), over a wide range of Zn contents ($y$) using field-dependent ac susceptibility (ACS) measurements. The ACS results enabled us to determine the superconducting transition temperature $T_c$, and the temperature $T_{gcp}$, at which the randomly oriented superconducting grains become coupled as a function of hole and disorder contents. We have analyzed the behavior of the GBs from the systematic evolution of the values of $T_{gcp}(p, y)$, $T_c(p, y)$, and from the contribution to the field-dependent ACS signal coming from the intergrain shielding current. Zn suppresses both $T_c$ and $T_{gcp}$ in a similar fashion. The hole content and the carrier localization due to Zn substitution seem to have significant effect on the coupling properties of the GBs. We have discussed the possible implications of these findings in detail in this article.

**PACS:** 74.72.Dn; 74.62.Dh; 74.25.Sv

**Keywords:** Zn doped La214; Critical current density; Grain boundary

\*Corresponding author. Email: salehnaqib@yahoo.com, Telephone: +88-0721-750288, Fax: +88-0721-750064

## 1. Introduction

From the point of view of large scale applications of cuprate superconductors, it is fair to say that critical current density, $J_c$, and the magnetic irreversibility field, $H_{irr}$, are the two most important parameters. Due to material specific problems, use of single crystals of high-$T_c$ cuprates is severely limited. Large sintered samples or thin films, on the other hand, often contain misaligned grain boundaries that weaken both the critical current density and the irreversibility field. Grain boundaries act as channels for the motion of Abrikosov or Josephson vortices, depending on the magnitude of the applied magnetic field [1, 2]. Impurities (intrinsic or extrinsic) act as pinning centers for magnetic flux lines that increases both $J_c$ and $H_{irr}$. Non-magnetic and iso-valent Zn substitution for in-plane Cu atoms diminishes the superconducting $T_c$ very effectively [3]. At the same time, Zn modifies the flux dynamics quite strongly and increases the pinning potential [4]. It has been found in the early days of cuprate superconductivity that Zn substitution leads to pair-breaking leading to a decrease in $T_c$ and charge carrier localization resulting in an increase in the resistivity [5, 6]. All these factors can affect the intergrain coupling and therefore the magnitudes of $J_c(T, H)$ and $H_{irr}(T)$. By controlling the grain boundary (GB) properties, the critical current density can be increased significantly [7]. Almost all high-$T_c$ cuprates have intrinsic disorders which affect the behavior at the GBs as well as that of the grains. In many ways such intrinsic disorders play a similar role as additional controlled disorder induced by electron irradiation or impurity atom substitutions in the $CuO_2$ planes [8, 9]. In this study we have investigated the coupling properties of the GBs of sintered $La_{2-x}Sr_xCu_{1-y}Zn_yO_4$ compounds over a wide range of Sr and Zn contents. From magnetic field dependent AC susceptibility (ACS) data the coupling temperature, $T_{gcp}$, of the GBs has been identified. Evolution of $T_{gcp}$ with Zn, and the magnetic field dependence of the ACS signal at low temperatures at different Sr concentrations, yielded valuable information regarding the quality of the GBs. Analysis of the ACS data showed that Zn affects both $T_c$ and $T_{gcp}$ in a similar fashion but the coupling properties vary significantly with hole content. The effect of Zn on the GBs also depend strongly on the hole concentration in the $CuO_2$ plane. We have discussed the possible implications of these findings in the subsequent sections.

## 2. Experimental Samples and results

Polycrystalline single-phase compounds of underdoped (UD) $La_{1.91}Sr_{0.09}Cu_{1-y}Zn_yO_4$ (with Zn contents $y$ = 0.00, 0.005, 0.01, and 0.024), slightly overdoped (SOD) $La_{1.81}Sr_{0.19}Cu_{1-y}Zn_yO_4$ (with $y$ = 0.00, 0.005, 0.01, 0.015, 0.02, and 0.024), and heavily overdoped (HOD) $La_{1.78}Sr_{0.22}Cu_{1-y}Zn_yO_4$ (with $y$ = 0.00, 0.005, 0.01) were prepared by solid-state reaction method using high-purity (% shown in brackets) $La_2O_3$ (99.999%), $SrCO_3$ (99.995%), CuO (99.9999%) and ZnO (99.999%) powders supplied by *Aldrich*. All these samples underwent the same heat treatment and the pellets were formed applying same pressure, consequently the densities of the sintered bars were almost identical (in the range from 6.09 to 6.20 gm/cm$^3$), irrespective of their composition. Samples were characterized by x-ray diffraction (XRD), room-temperature thermopower ($S$[290 K]), and low-field ($H_{rms}$ = 1 Oe, $f$ = 333.33 Hz) AC susceptibility (ACS) measurements. Details of sample preparation and characterization can be found in ref. [10]. Representative XRD profiles are shown in Fig. 1. From XRD no trace of impurity phase was found in any of the compounds used in this study. The values of $S$[290 K] remained almost unchanged with Zn substitution, which indicated that the hole content, $p$ ($\equiv x$), in the $CuO_2$ plane was not affected by iso-valent Zn substitution [11]. $T_c(x, y)$ values were determined from the diamagnetic onset of the low-field (1 Oe) ACS signal (see Figs. 2). At a given $x$, $T_c(y)$ values gave an independent idea about the level of Zn substituted in the Cu sites. The GB coupling temperature has been defined as the temperature at which the intergrain shielding current starts to contribute to the field dependent ACS signal. All the ACS results presented in this study were obtained in the field-cooled mode. This enabled us to discuss the field, disorder (Zn), and $p$-dependent ACS data in terms of the superconducting volume fraction [12], when appropriate. Examples of obtaining $T_{gcp}$ are shown in Figs. 3. At temperatures just below $T_c$, all the grains are decoupled as superconductivity in the GB regions are weak and the superconducting (SC) order parameter inside individual grains are small. The ACS signal in this region consists of contribution only from individual grains and as long as the magnetic field is weak this signal varies linearly with the field. Depending on the quality of the GBs, the field dependent ACS signal increases at lower temperatures largely because the intergrain shielding currents start to contribute.

Comparison of the ACS signals of powdered compound to that of a sintered one gives an estimation of the intergrain contribution, since grains are isolated in the powdered sample. We show such a plot in Fig. 4 (with $x = 0.09$, $y = 0.00$). The average radius of the particulates in the powdered compound was ~ 2.0 µm. In comparison to the sintered one, the ACS signal from the powdered compound is drastically reduced due to the absence of the intergrain shielding current. The gradual $T$-dependence of the ACS signal for powders is due to the decrease in the magnetic flux penetration depth as temperature is lowered below $T_c$ [13]. We have also shown the field dependent ACS data for the same powdered compound in Fig. 5. The field-normalized ACS signals show both quantitatively and qualitatively identical behavior, giving a firm indication that the non-linear field dependent ACS signals below $T_{gcp}$ shown in Figs. 3, originate from the GBs.

## 3. Analysis of the field dependent ac susceptibility data

The effect of Zn content on the parameter defined by $(T_c - T_{gcp})$ at different hole content is shown in Fig. 6. This parameter is indicative of how fast the intergrain current comes into play as temperature is lowered below $T_c$. Somewhat surprisingly, no systematic hole or disorder content dependence can be found in $(T_c - T_{gcp})$. This may imply that the initial coupling between the grains is determined only by the compactness (density) of the sintered compounds, irrespective of the level of hole or impurity. Next, we have investigated the role of the hole and disorder contents on the GB by comparing the field dependence of the ACS signal at fixed levels of Zn contents at 4.2 K, as shown in Fig. 7. For the sintered compounds the magnitude of the ACS signal increases non-linearly with increasing magnetic field (as the GBs become more and more decoupled at higher magnetic fields), as a result the field normalized ACS signal (expressed in the units of ACS signal/gm-Oe) decreases with increasing field irrespective of the hole concentration in the $CuO_2$ plane (see Fig. 3). The degree of this decrement depends strongly on the hole content. This result is expected because as $x$ increases the sample becomes more conducting and so do the GBs. Consequently, the GB connectivity improves and larger fields are required to decouple them. It is interesting to note that the highest ACS signal is obtained for the $x = 0.19$ compounds at all fields and at all levels of Zn substitution, even though the electrical conductivity of the $x =$ 0.22 compound is always higher at a fixed value of $y$. The answer to this may lie in the facts that at $x = 0.19$ the superfluid density and the SC condensation energy are at their maximum [14]. This should lead to a stronger proximity effect and consequently to a stronger intergrain coupling. For the UD compounds the effect of magnetic field is small for the Zn substituted ones (see Fig. 7c). In Fig. 7b where Zn content is smaller, the field insensitivity appears at higher applied magnetic fields. This implies that even though the $T_c$ of the $x = 0.09$ compound is higher than that of the $x = 0.22$ one, Zn degrades superconductivity much more efficiently in these UD compounds. As superconductivity is severely weakened due to Zn inside the grains of the UD compounds, these grains are almost completely decoupled even when a small magnetic field (*e.g.*, 3 Oe) is applied. Therefore, the ACS signal at higher magnetic fields (*e.g.*, 10 Oe) consists of only the intra-grain contribution, which remains quite insensitive in the range of magnetic fields used in this study. Such saturation of the ACS signal with magnetic field also takes place for the OD compounds but at higher Zn concentrations (data not shown in this paper). For example, the low-$T$ (4.2 K) field normalized ACS signals obtained with magnetic fields of 3 Oe and 10 Oe are almost identical for the 2.4% Zn substituted $La_{1.81}Sr_{0.19}Cu_{1-y}Zn_yO_4$ compound. It is possible to estimate roughly the evolution of the superconducting volume fraction [12] at low-$T$ with magnetic field, hole content, and Zn concentration from the ACS data. We have analyzed this in Figs. 8, where the ACS signal has been normalized with the signal for the Zn-free compound at 4.2 K. Zn reduces the SC volume fraction at all magnetic fields and this reduction is highly $p$ dependent. For example, at 1 Oe, the normalized ACS signal is reduced to 24%, 93%, and 89% for the $x = 0.09$, 0.19, and 0.22 compounds, respectively, when $y = 0.01$. Once again a drastic fractional reduction in ACS signal is found for the UD compound compared to the HOD one with a lower $T_c$. We believe this is due to the following factors (a) the superfluid density and the SC condensation energy are low in the UD compound and (b) there is a substantial pseudogap at low energy in the quasi-particle (QP) spectrum, which enables Zn to break the Cooper pairs very effectively [5, 15]. Similar conclusions were reached in earlier studies [16, 17]. Zn is least effective in reducing the SC volume fraction (see Figs. 8a and 8b) for the $x = 0.19$ compound, giving further indication that the superconductivity is at its strongest at this hole content. Apparently, the behavior shown in Fig. 8c is somewhat different. This is largely because the signal from the $x = 0.09$ compound is very weak due to significant decoupling of the GBs at 10 Oe and the role of Zn is largely to weaken superconductivity only inside the grains.

**4. Discussions and Conclusion**

It should be noted that, both $T_c(x, y)$ and $T_{gcp}(x, y)$ have been identified at a magnetic field of 1 Oe. This does not affect the various observations mentioned in section 3 in any significant way. We have found that irrespective of the hole content, Zn always has a degrading effect on the coupling of the GBs. The GB connectivity shows a complex behavior. It does not dependent only on the conductivity of the compound but also on the superfluid density and superconducting condensation energy inside the grains. Thus for the UD compounds ACS signal decreases quite rapidly as magnetic field or Zn content is increased, even though $T_c$ of the pure sample is quite high. Among all the samples used in this study, the best behavior is observed for the $x = 0.19$ compounds. This agrees completely earlier studies [16, 17], where the possible role of the pseudogap in the QP spectrum on the irreversibility field and critical current density were studied in detail. The PG seems to vanish at this particular doping [14].

This paper deals with the behavior of the GBs in high-$T_c$ sintered cuprates as a function of in-plane hole and disorder contents. From the analysis of the field-cooled ACS data for samples with wide range of compositions, the followings conclusions can be reached (1) the temperature at which the intergrain coupling starts (with respect to $T_c$) is quite insensitive to the composition of the sample (as shown in Fig. 6), (2) the quality of the GBs, and therefore the magnitude of the intergrain shielding current, depends strongly on the hole and Zn contents – grains are weakly coupled in the UD compound and Zn degrades GBs quite severely in this region. The effect of Zn on the GBs is less severe for the OD compounds, (3) strength of intergrain coupling seems to be determined by the SC condensation energy, superfluid density, and grain conductivity, (4) the strongest coupling between the GBs is observed for the $x = 0.19$ compounds because the superconducting condensation energy and the superfluid density are at their maximum at this composition [14], and (5) PG plays a significant role in determining the pair-breaking strength due to Zn [5, 15], therefore Zn breaks Cooper pairs very effectively inside individual grains and reduces the superfluid density in the UD region where PG is large; consequently intergrain coupling becomes very weak. In the OD compounds, pair-breaking scattering rate is reduced due to the absence of the PG in the QP electronic density of states [5, 15], leading to better coupling between the GBs. We would also like to mention here that some earlier studies [4], on the magnetic flux dynamics of cuprate superconductors, have found a Zn induced enhancement in the pinning potential for vortices. This study, on the other hand, indicates that even if Zn acts as pinning centre, the degrading effects due to pair-breaking and carrier localization will outweigh significantly any such positive contribution to the critical current density or irreversibility field in sintered compounds. The remarks made above could also be true for any other type of intentional or un-intentional in-plane disorder capable of SC pair-breaking.

## Acknowledgements

The authors acknowledge the Commonwealth Commission, UK, and Trinity College, University of Cambridge, UK, for financial support.

**Figure captions**

Figure 1: XRD profiles of representative $La_{1.91}Sr_{0.09}Cu_{1-y}Zn_yO_4$ compounds. The Miller indices for various diffraction peaks are shown in Fig. 1a. The Zn contents are given in the plots.

Figure 2 (Color online): Low-field (1 Oe) ACS signal of representative $La_{2-x}Sr_xCu_{1-y}Zn_yO_4$ sintered compounds with (a) $x = 0.09$ (UD) and (b) $x = 0.22$ (HOD). Sample compositions are shown in the plots. The arrows mark the onset of diamagnetic transition and therefore $T_c$ at 1 Oe. The horizontal dashed straight line shows the trend of the normal state ACS signal.

Figure 3 (Color online): Locating the grain coupling temperature, $T_{gcp}$, for the sintered $La_{1.81}Sr_{0.19}Cu_{1-y}Zn_yO_4$ compounds with (a) $y = 0.00$, (b) $y = 0.005$, and (c) $y = 0.01$.

Figure 4 (Color online): ACS signal with (sintered) and without (powdered) intergrain contribution at 1 Oe for the $La_{1.91}Sr_{0.09}CuO_4$ compound.

Figure 5 (Color online): Field-normalized ACS signals for the powdered $La_{1.91}Sr_{0.09}CuO_4$ compound. Magnetic field independence of the ACS signal for powders shows that contributions from the GBs are absent here.

Figure 6 (Color online): $(T_c - T_{gcp})$ versus Zn content for the $La_{2-x}Sr_xCu_{1-y}Zn_yO_4$ compounds. The Sr contents are given in the plot.

Figure 7 (Color online): Field normalized ACS signal versus magnetic field for the $La_{2-x}Sr_xCu_{1-y}Zn_yO_4$ compounds at 4.2 K, with (a) $y = 0.00$, (b) $y = 0.005$, and (c) $y = 0.01$. The dashed lines are drawn as guides for the eye.

Figure 8 (Color online): Normalized ACS signal (see text for details) versus Zn content for $La_{2-x}Sr_xCu_{1-y}Zn_yO_4$ compounds at 4.2 K, with applied magnetic fields ($H_{rms}$) of (a) 1 Oe, (b) 3 Oe, and (c) 10 Oe. The dashed lines are drawn as guides for the eye.

Figure 1

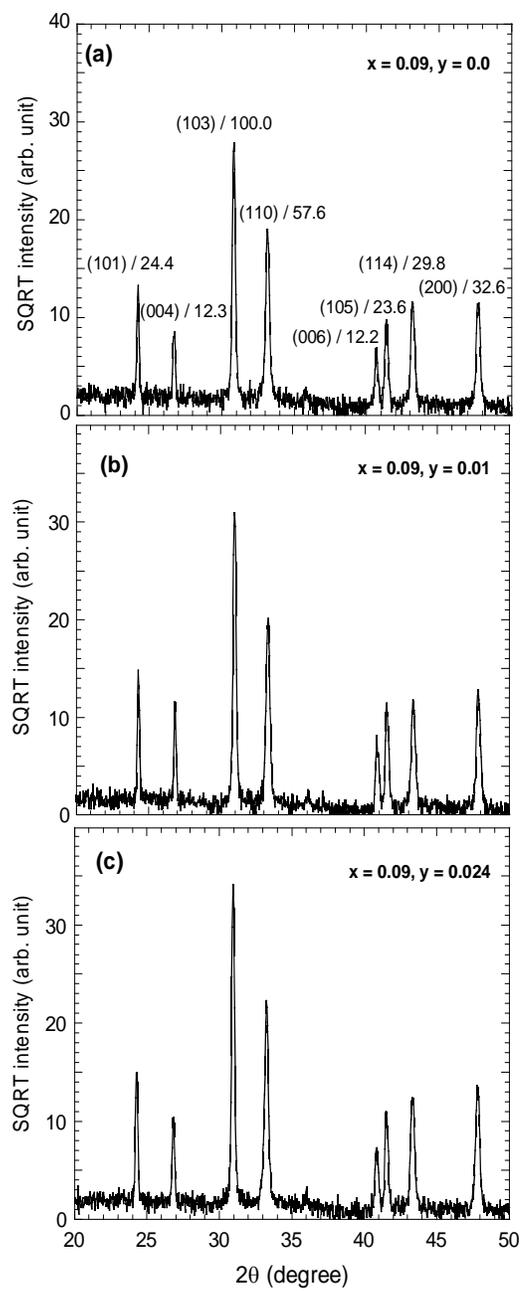

Figure 2

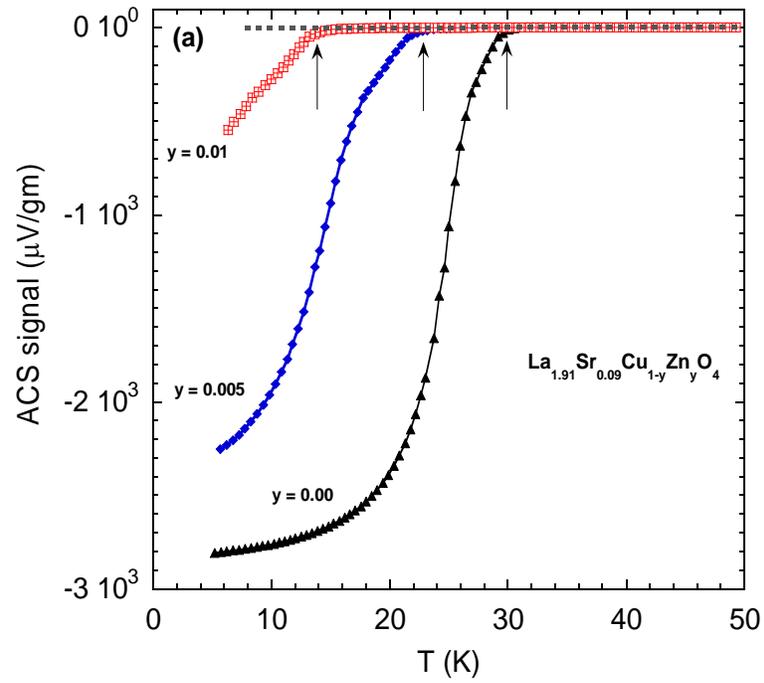

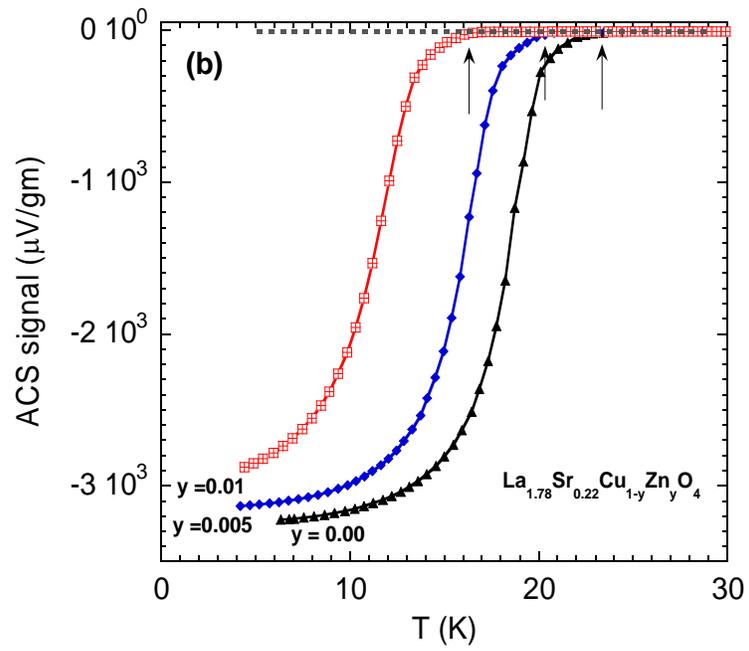

Figure 3

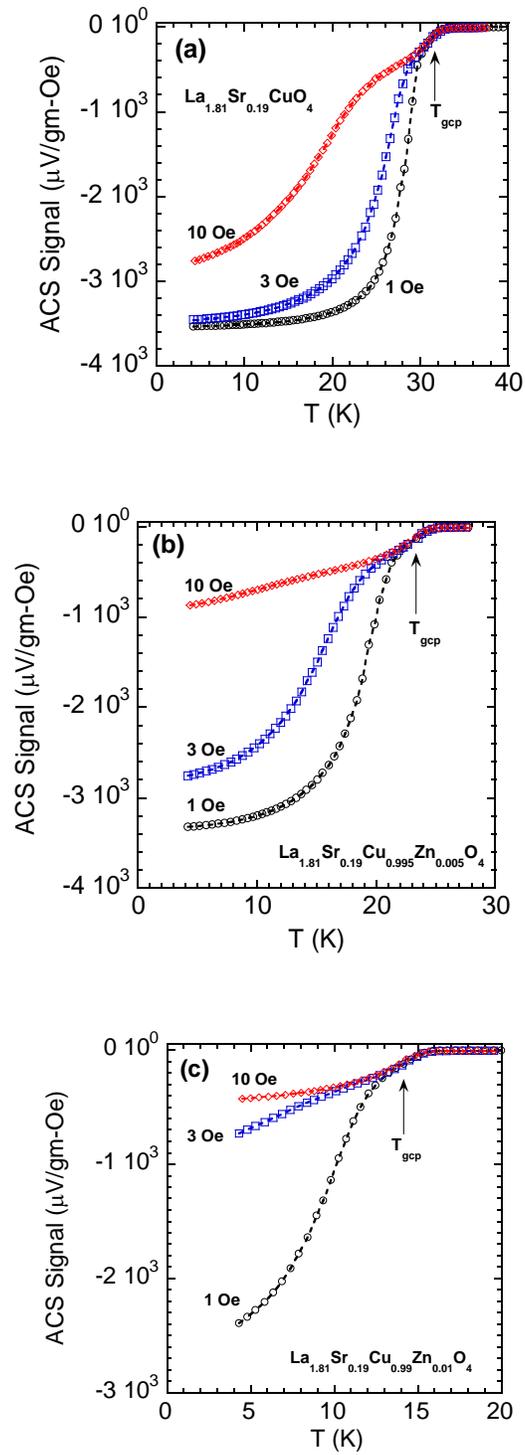

Figure 4

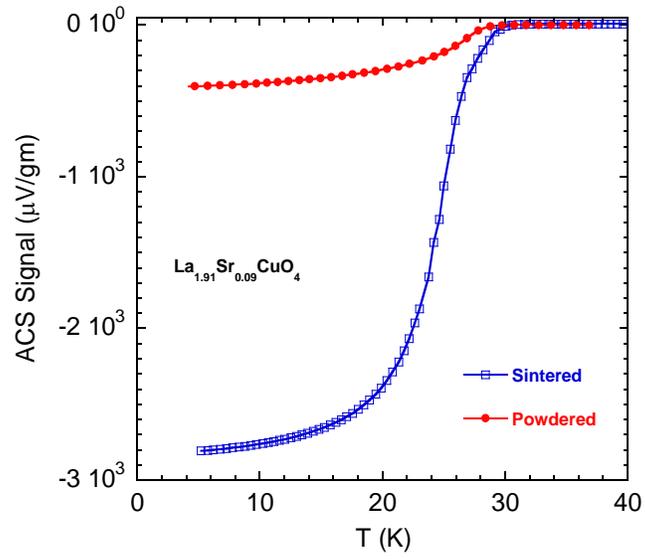

Figure 5

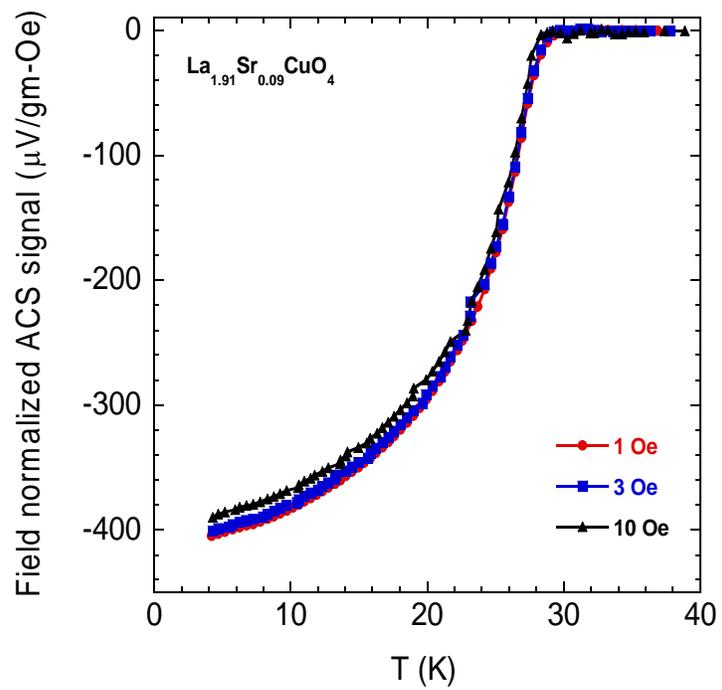

Figure 6

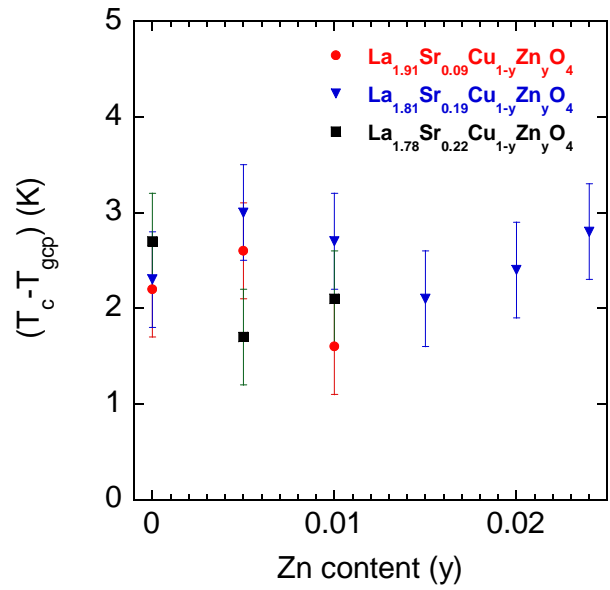

Figure 7

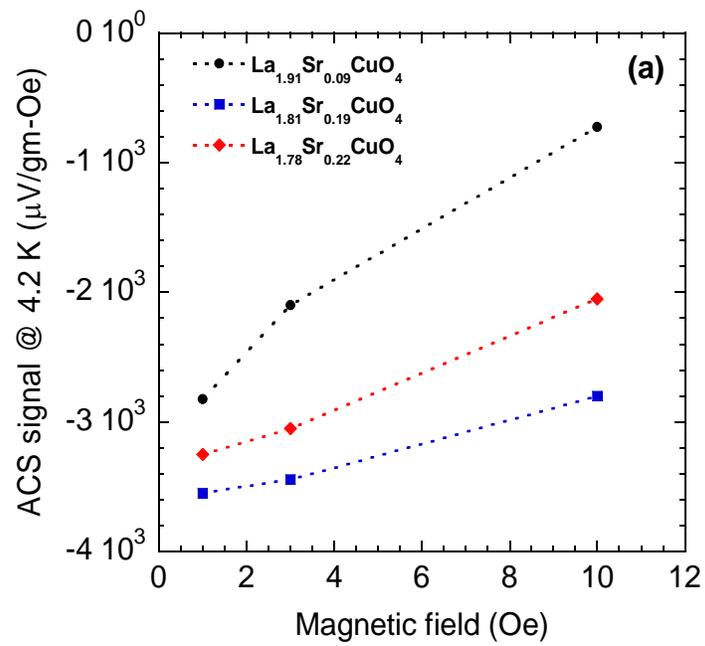

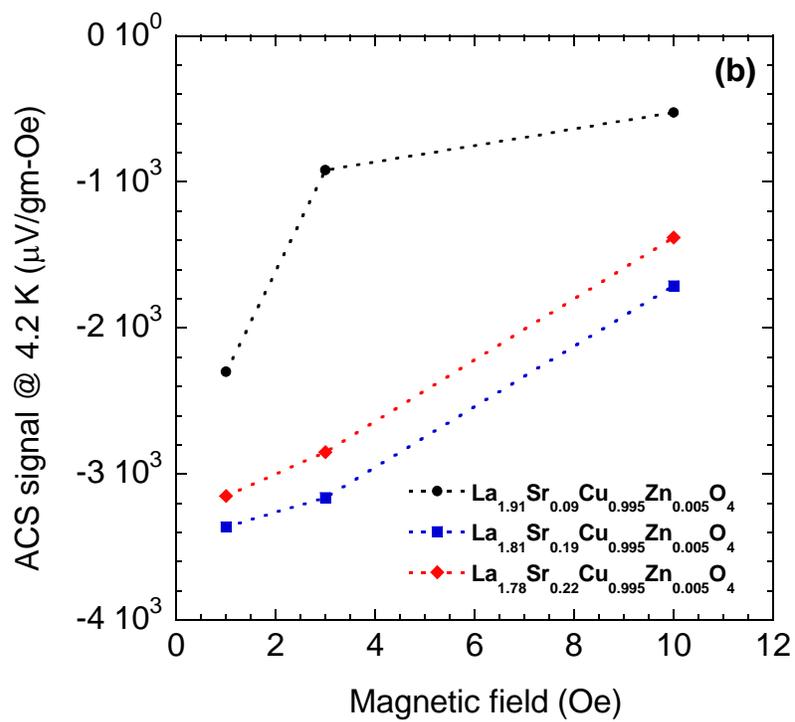

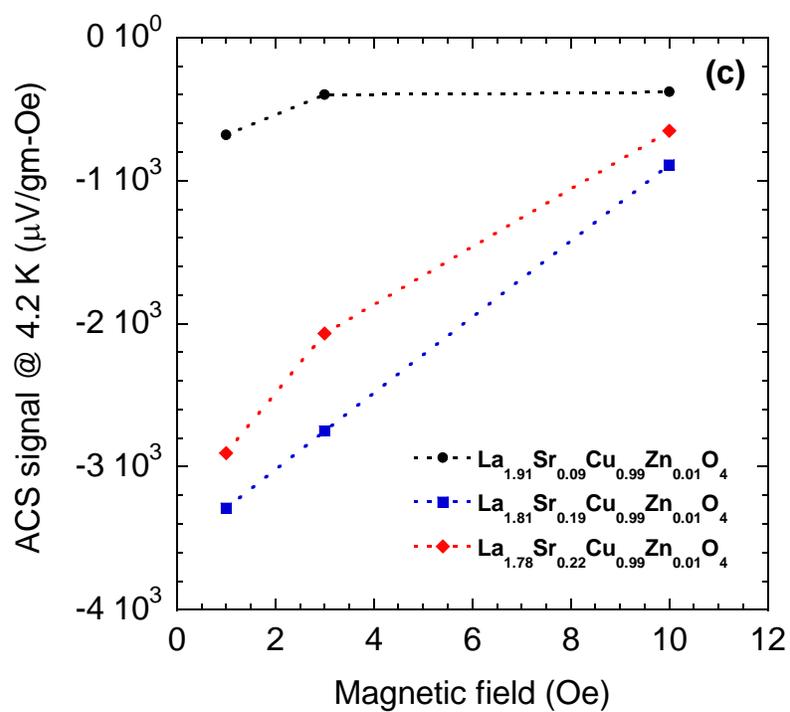

Figure 8

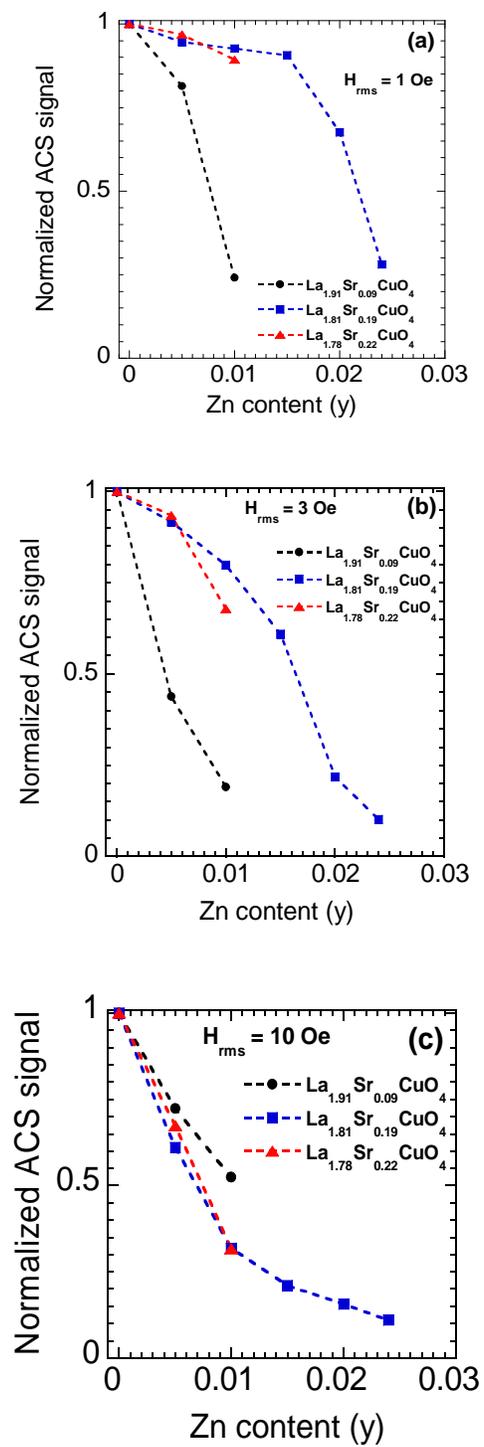